\documentclass[]{aa}
\usepackage{graphicx}
\usepackage{natbib}
\usepackage{txfonts}
\def\mh{\,$\mu$Hz}

\def\co{CoRoT}

\def\lg{\ensuremath{\log g}}
\def\hd{HD\,49933}
\def\p1{Paper\,I}

\begin{document}
   \title{The nature of p-modes and granulation in HD\,49933 observed by CoRoT
   \thanks{The \co\  space mission was developed and is operated by the French space agency CNES, with participation of ESA's RSSD and Science Pograms, Austria, Belgium, Brazil, Germany, and Spain}}

   \author{Thomas Kallinger\inst{1, 2}
   	  \and
          Michael Gruberbauer\inst{1, 3}
          \and
          David B. Guenther\inst{3}
          \and
          Luca Fossati\inst{1}
          \and
          Werner W. Weiss\inst{1}
            }

   \offprints{kallinger@phas.ubc.ca}

   \institute{
   Institute for Astronomy (IfA), University of Vienna, T\"urkenschanzstrasse 17, A-1180 Vienna
              \and
  Department of Physics and Astronomy, University of British Columbia, 6224 Agricultural Road, Vancouver, BC V6T 1Z1, Canada
		\and 
              Institute for Computational Astrophysics, Department of Astronomy and Physics, Saint Marys University, Halifax, NS B3H 3C3, Canada
             }

   \date{Received ; accepted }

\abstract
{Recent observations of \hd\ by the space-photometric mission \co\ provide \emph{photometric} evidence of solar type oscillations in a star other than our Sun. The first published reduction, analysis, and interpretation of the \co\ data yielded a spectrum of p--modes with $l$ = 0, 1, and 2.}
{We present our own analysis of the \co\ data in an attempt to compare the detected pulsation modes with eigenfrequencies of models that are consistent with the observed luminosity and surface temperature. }
{We used the Gruberbauer et al. frequency set derived based on a more conservative Bayesian analysis with ignorance priors and fit models from a dense grid of model spectra. We also introduce a Bayesian approach to searching and quantifying the best model fits to the observed oscillation spectra.}
{We identify 26 frequencies as radial and dipolar modes. Our best fitting model has solar composition and coincides within the error box with the spectroscopically determined position of \hd\ in the H-R diagram. We also show that lower-than-solar Z models have a lower probability of matching the observations than the solar metallicity models. To quantify the effect of the deficiencies in modeling the stellar surface layers in our analysis, we compare adiabatic and nonadiabatic model fits and find that the latter reproduces the observed frequencies better.}
{}

\authorrunning{Kallinger et al.}
\titlerunning{The nature of p-modes and granulation in HD\,49933}
   \maketitle

\section{Introduction}	\label{sec:intro}
\co\  (Convection, Rotation, and planetary Transits) is a space mission focusing on asteroseismology and the detection of exoplanetary transits. See \citet{boi06} for an  overview of the technical details and \citet{bag06} for a review of the asteroseismology aspects of the mission. 
\hd\ (F5\,V, m$_V$\,=\,5.77) is a main sequence star with an effective temperature of about 6500\,K,  \lg\ $\approx 4.0$, L/L\sun\ $\approx 3.6$, and [Fe/H]\,=\,-0.38. The determination of these values is described in our Sect.\,\ref{sec:fun}. \hd\ is similar to Procyon, except with a lower metal abundance. \hd\ was observed during the initial run of  \co\  from February 6 to April 7, 2007, only little more than one month after launch of \co\ on December 27, 2006. 

The reduction of the \co\ photometry to the N2 data format is described in \citet{apo08}. They also provide the first frequency determination, a preliminary echelle diagram, and a mode identification for \hd. Because their conclusions differ from ours we review their procedure here highlighting areas where we have adopted a different set of premises. They use a maximum likelihood estimator technique to fit Lorentzian mode profiles to the observed power spectrum with symmetric rotational splitting components included with the non-radial modes.  In order to reduce the number of free parameters, they assume the same line width for modes of the same radial order, and they assume the mode heights are related to each other by fixed ratios. For the rotational splitting components, they assume the splitting is symmetric in frequency with the split-mode height ratios determined from the inclination angle parameter.

Our approach to identifying frequencies \citep[][hereafter \p1]{gru08} is distinct, in that we emphasize the \emph{reliability} of the frequency detections over the  \emph{quantity} of detected frequencies. Without prejudging the identity of a mode, we first simply ask if a mode is statistically visible in the data. Then when we have ascertained its existence above some statistical measure, we try to fit stellar models to the resultant list of frequencies. When comparing observed frequencies with stellar model frequencies, even a few incorrectly identified frequencies can significantly complicate the analysis or even lead to a misinterpretation. 

In \p1\ we describe the fitting of solar-type p-modes as a parameter estimation problem without the need to give preference to any specific model. A Markov-Chain, Monte Carlo technique for a Bayesian analysis of Lorentzian profiles was used rather than fitting complex models to the observations. Specifically, we did not make any assumptions about mode heights, knowing that noise in the data, itself, can affect the observed heights of individual modes. Rather, we used our fits to the mode heights compared to the noise level to characterize the viability of the mode being detected. 
We, therefore, only assumed the Lorentzian profiles have the same line width. 
This assumption might distort the correct determination of the mode height and line width parameters, because it is known, e.g., for the Sun \citep[e.g.][]{cha05}, that the mode lifetime is a function of frequency. To fix the mode line widths to a single parameter, however, should not (or only marginally) influence the determination of the mode frequencies, which is the parameter we are most interested in.
We extracted a number of radial and $l$ = 1 modes. We found no evidence in the \co\ data of statistically signifiant $l$ = 2 modes or for rotationally split components. If they exist, of course, these components are below the threshold of our mode-height-detection criterion. We argue that the \citet{apo08} approach is too restrictive in that it presumes specific characteristics of the modes identity. At this early stage in the analysis of such stars, we are not comfortable making any assumptions about the modes even at the expense of identifying fewer modes. 

In Sect.\,\ref{sec:fun} we re-examine the fundamental parameter determinations of \hd . In Sect.\,\ref{sec:mod} we describe two methods for searching large grids of stellar models for a best fit between observed and model frequencies. One of the methods, again a Bayesian approach, allows us to estimate the statistical likelihood of the fit compared to other models in our grids. Our best fitting model has solar composition and is located  in the H-R diagram within the spectroscopically determined error box of  \hd.

\section{Fundamental parameters} 		\label{sec:fun}

Eight different values for the effective temperature, based on spectroscopy and photometry and ranging from 6467\,K to 6780\,K are cited by \cite{gil06}. \cite{bla98} determined the effective temperature to be 6512\,K by using the IR flux method.  In their paper on the CoRoT asteroseismic observations of HD\,49933, \cite{apo08} adopt an effective temperature of 6780$\pm130$\,K, which originally comes from \cite{bruntt08}. 
Another frequently cited value for the effective temperature is 6576$\pm$98\,K, which is derived from Stroemgren photometry using the \citet{moon1985} calibration. This procedure also provides a surface gravity of \ensuremath{\log g} =  4.30$\pm$0.09. Many of the published temperature values are based on standard photometric calibrations or automatic spectroscopic determinations. As a consequence of the large scatter of the published values for this essential stellar parameter we performed our own temperature determination using high-quality spectroscopic observations.

In February 2006, 10 high-resolution spectra (R$\sim$115\,000) were obtained for HD\,49933 with the HARPS spectrograph at the ESO-LaSilla 3.6\,m telescope. We retrieved these spectra from the ESO archive and increased the signal--to--noise ratio ($\sim$500 in a 0.5\,\AA\ continuum bin around 5000\,\AA ) by co-adding. The continuum normalization was done with a low order polynomial fit to regions free of visible spectral lines.

\begin{figure}[ht]
	\begin{center}
	\includegraphics[width=0.5\textwidth]{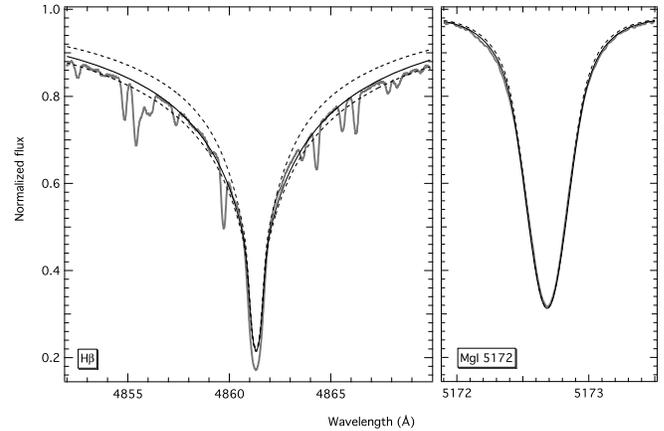}
	\caption{Observed H$\beta$ and \ion{Mg}{i} line profile of HD\,49933 (gray line), compared to synthetic profiles. The thin black lines correspond to synthetic line profiles assuming $T_{\mathrm{eff}}$ and \ensuremath{\log g} to be 6450\,K and 3.9, respectively.
	\emph{Left panel}:  The dashed lines shows the synthetic profile by increasing (lower line) and decreasing (upper line) $T_{\mathrm{eff}}$ by 50\,K. For plotting reasons we excluded the metallic lines from the synthetic hydrogen line profiles.
\emph{Right panel}:  The dashed black line shows the synthetic profile by increasing \ensuremath{\log g} by 0.1\,dex.	
} 
	\label{hbeta} 
	\end{center} 
\end{figure}

To derive a consistent and reliable value for $T_{\mathrm{eff}}$ we compared the observed hydrogen line profiles with synthetic spectra computed with {\sc Synth3} \citep{synth3}. In the given temperature range the hydrogen lines are very sensitive to temperature variations, are less sensitive to \ensuremath{\log g} variations, and depend on the metallicity of the atmosphere. We calculated model atmospheres with {\sc LLmodels} \citep{denis2004}, an LTE code that uses direct sampling of the line opacities and allows models to  be computed with an individualized abundance pattern. \ion{Mg}{i} line wings \citep[$\lambda\lambda$ 5167 5172, 5183\,\AA;][]{vdw} and the ionization equilibrium of several additional elements were used to determine the gravity. We started our model atmosphere iteration with the most convincing value of $T_{\mathrm{eff}}$, based on models with solar abundances, and a surface gravity derived from photometry. We derived the abundances of several elements from equivalent widths and then determined  \ensuremath{\log g} from a set of models using the previously obtained $T_{\mathrm{eff}}$ and abundances. With this value for the gravity we determined anew the abundances and used them to improve $T_{\mathrm{eff}}$. 
After two iterations we converged to $T_{\mathrm{eff}}$ = 6500$\pm$50\,K and \ensuremath{\log g} = 4.0$\pm$0.15, which is significantly cooler than the value adopted by \cite{apo08}. The small error in $T_{\mathrm{eff}}$ can be explained by the high temperature sensitivity of the hydrogen line profiles (see left panel of Fig.\,\ref{hbeta}). Regardless, for the following considerations we adopt a more conservative error for the effective temperature of $\pm$75\,K. 
The atomic line parameters were taken from the VALD database \citep{vald1,vald2,vald3}, except for the Van der Waals broadening constants, which were taken from \citet{vdw}. The quoted error for \ensuremath{\log g} is similar to other stars with comparable effective temperatures, such as Procyon \citep{vdw}, and using the same method. 

It is useful in asteroseismic modeling to know the metallicity of the star. According to our spectroscopic analysis the atmospheric metallicity is Z = 0.008$\pm$0.002 where iron is under abundant compared to the Sun \citep{grev96}, in agreement with published values. Carbon and oxygen, on the other hand, have near solar abundances. This variation could be due to the effects of gravitational settling and radiative levitation in the thin convective envelope of the star. The true interior abundance, relevant for stellar models and pulsation analysis, is uncertain. Ideally, stellar models should include the effects of element diffusion. 
Further details of the spectroscopic analysis with a discussion of the literature can be found in \cite{rayab09}.

Using the Hipparcos parallax $\pi = 33.69\pm0.42$\,mas \citep{van07} we abtain the absolute visual magnitude M$_\mathrm{V} = 3.418\pm0.026$. Interpolating the tables of \cite{lej01} we obtain a bolometric correction BC$_\mathrm{V} = -0.054\pm0.005$ (for [Fe/H] = -0.4), which is more than twice the value used by \cite{apo08}. With M$_\mathrm{bol, \sun}$ = 4.75 we obtain L/L\sun = 3.58$\pm$0.10 (Appourchaux et al. 2008 use 3.39$\pm$0.08) and a corresponding radius of R\,=\,1.46$\pm$0.05 R\sun.

\section{Asteroseismic analysis}		\label{sec:mod}

For our analysis we use the frequencies given in Paper I (see Tab.\,5 therein) where we exclude P25 and P26 following the suggestion of the authors. They did not assign credibility to these values arguing with the ambiguity in their marginal frequency distribution, which is apparent in the listed large uncertainties. 
In Paper I, no specific mode identifications to the frequencies were assumed and mode heights were left unconstrained. \cite{apo08}, on the other hand, assumed specific mode identifications ($l$ = 0, 1, and 2) and fixed relative mode heights. 

We search a dense and extensive grid of stellar model oscillation spectra looking for the model spectra that matches best the observed frequencies to within the known uncertainties of the observed frequencies. We use a simple $\chi^2$ formula to quantify how well the spectra match each other. This approach is fast and efficient and can handle sparse and contaminated oscillation spectra. Although a lower $\chi^2$ does indicate a higher likelihood, the computed $\chi^2$s cannot be used to associate a confidence level to the inferred parameters. 
We, therefore, introduce a new method based on a Bayesian approach that uses the error distributions of the observed frequencies to quantify in terms of a probability how well the frequencies of a given model match the observations. 

The grids of models were constructed using the Yale Stellar Evolution Code \citep[YREC;][]{guenther92}. The grids include models with masses ranging from 0.6 to 2.00\,M\sun\ in steps of 0.005\,M\sun\ and from 2.00 to 5.00\,M\sun\ in steps of 0.01\,M\sun . Each evolutionary track runs from the zero age main sequence to halfway up the giant branch. The solar grid is based on a near solar composition and calibrated mixing length parameter (Z = 0.02, Y = 0.27, $\alpha$=1.8). The low Z grid, which has the same mass resolution but is not nearly as extensive in its coverage, is based on the same mixing length parameter and (Z, Y) = (0.008, 0.24).

The constitutive physics of the models include OPAL98 \citep{igl96}, the \citet{ale94} opacity tables, and the Lawrence Livermore National Laboratory equation of state tables \citep{rog86,rog96}. The standard B\"ohm-Vitense \citep{vit53,vit58} mixing length theory is used to model convection. The mixing length parameter used to describe the temperature gradient in convective regions was adjusted from calibrated solar models and rounded to two significant digits. We have assumed standard solar mixture \citep{grev96}. Although the effects of helium and heavy element diffusion are included in the model physics, diffusion is automatically turned off shortly after turn-off to avoid known computational problems associated with very thin convection zones. Each model was resolved into approximately 2000 shells, with two-thirds of the shells placed in the envelope and atmosphere. Guenthers nonradial nonadiabatic stellar pulsation program \citep{gue94} was used to compute the adiabatic and nonadiabatic pulsation spectra of each of the models in the grid. The nonabatic code includes the effects of radiation on the modes but does not include the, probably equally important, effects of convection. 

\subsection{$\chi^2$ mode matching}	\label{ssec:match}

We begin our model analysis using the $\chi^2$ mode matching method first introduced by \cite{gue04}. The method is basic. Large grids of stellar models and their spectra are constructed. The stellar model spectra are compared one-by-one to the observed spectrum. How well the two spectra compare is quantified by the following $\chi^2$: 
	\begin{equation}
	\chi^2 = \frac{1}{N_o} \sum_{i=1}^{N_o} \frac{(\nu_{m,i} - \nu_{o,i})^2}{\sigma_{o,i}^2},
	\label{eq:chi}
	\end{equation}
where $\nu_{o,i}$ and $\nu_{m,i}$ are the observed and corresponding model eigenfrequency of the $i$-th mode, respectively. The observational uncertainty is given by $\sigma_{o}$, and $N_o$ corresponds to the total number of modes used for the fit. This approach provides a good estimate of how well a set of observed frequencies coincides with a model eigenspectrum. A value of $\chi^2 \leq$ 1.0 means that, on average, the model agrees within the uncertainties of the observed frequencies.  In other words, the $\chi^2$ provides a measure for the average normalized deviation between observed and model frequencies and should not be misinterpreted in terms of a probability. For more details, see, e.g., \citet{gue05} or \cite{kal08b}. 

To find a best model fit to the 26 observed frequencies (Tab.\,5 in Paper I) we searched for modes from $l$ = 0 to 3, inclusive. In our range of interest in the H-R diagram the ``resolution" of our grid, i.e., the step in frequency for a mode of given degree and radial order from model to model, is about 4\mh\ ($\sim$0.2\% of the observed frequency with the highest amplitude). This is quite high compared to the uncertainties in the observations. We therefore interpolated our grid linearly along the evolutionary tracks, effectively increasing the grid density by a factor of 10. 

\begin{figure}[ht]
   \centering
      \includegraphics[width=0.48\textwidth]{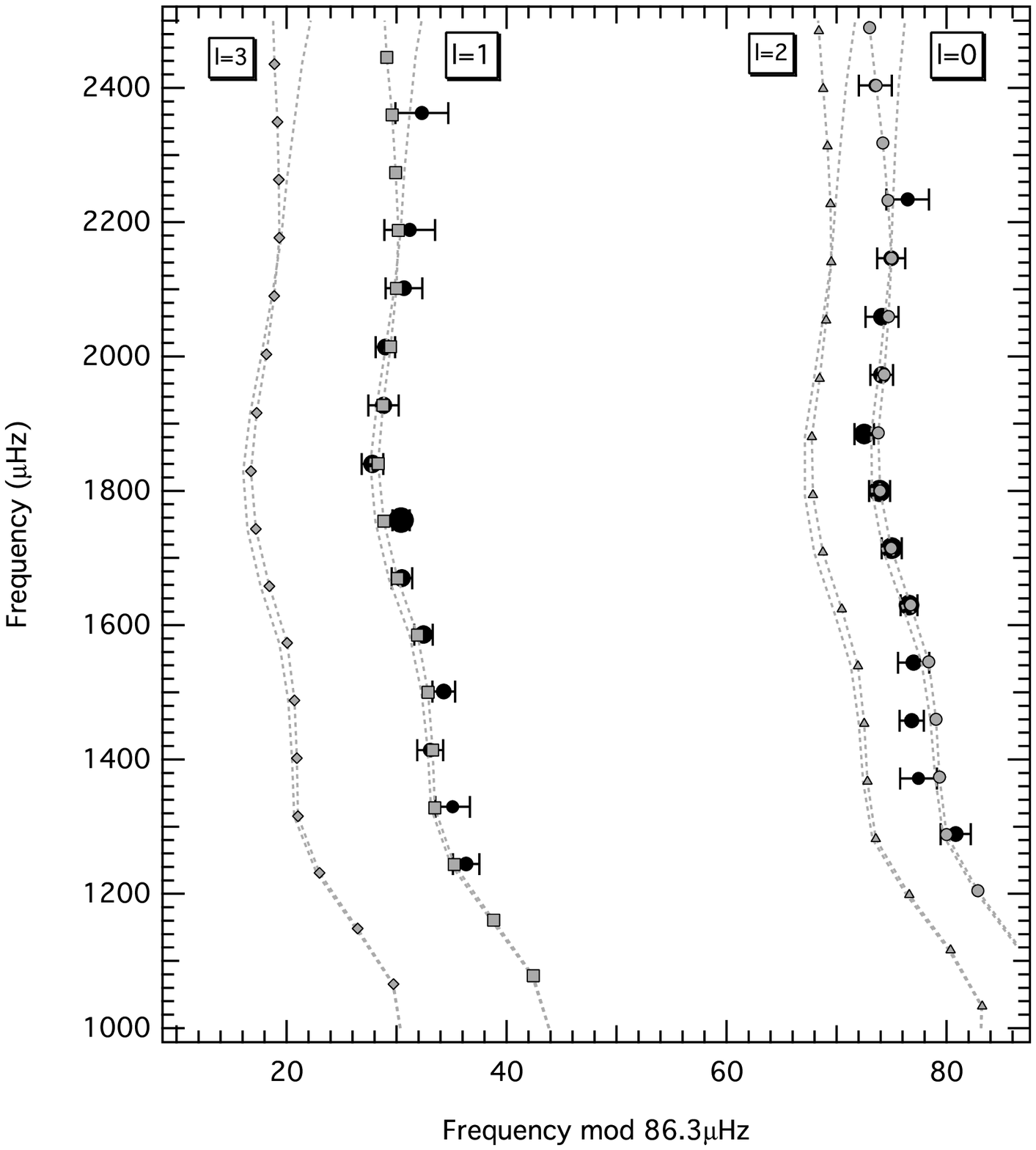}
      \includegraphics[width=0.48\textwidth]{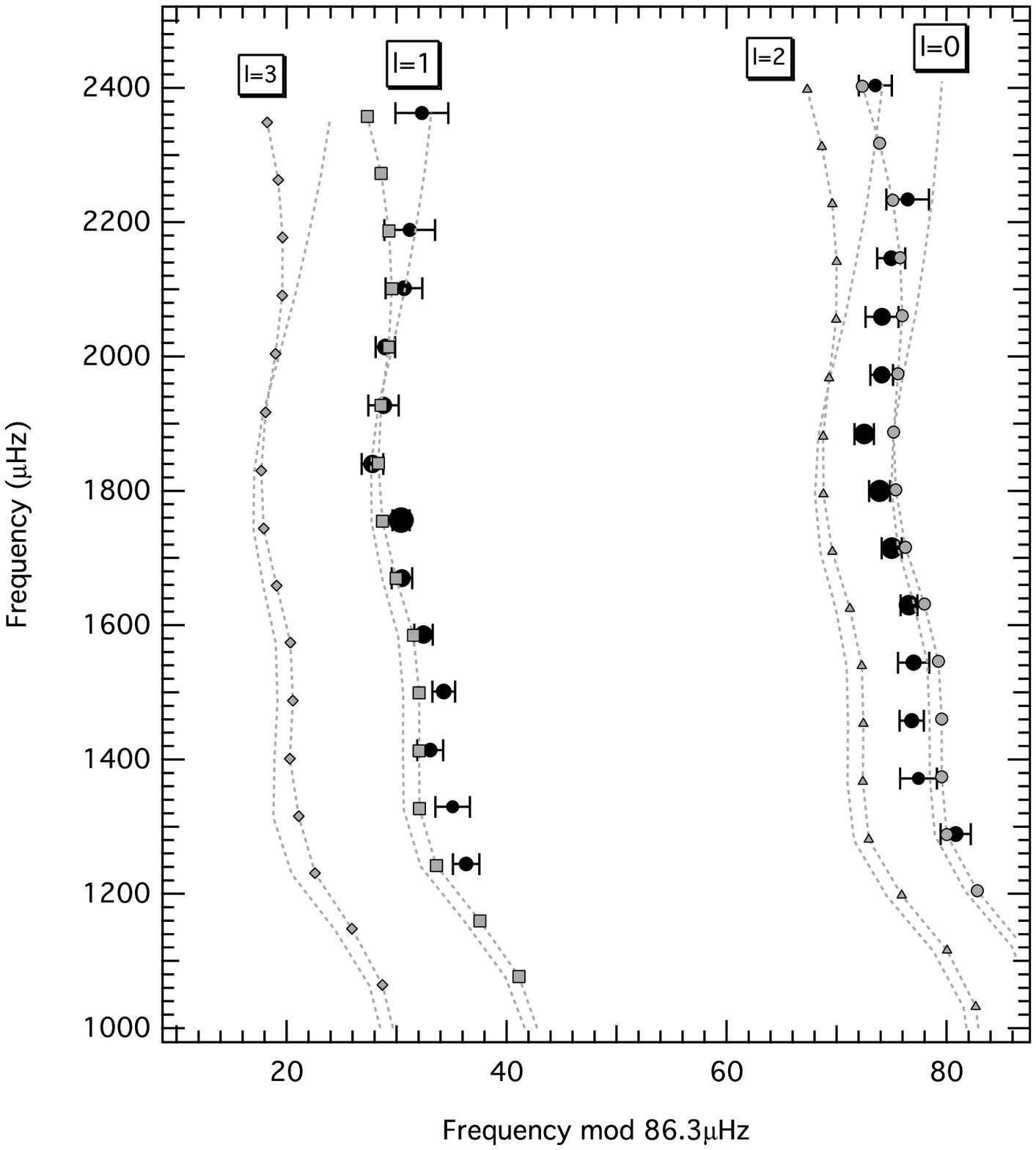}
      \caption{\emph{Top panel}: Echelle diagram of the observed oscillation frequencies and the adiabatic and nonadiabatic eigenfrequencies of the best fitting model from the solar abundance model grid. The observed frequencies are shown as filled circles with the error bars indicating the observational uncertainties. The nonadiabatic model frequencies used in the analysis are displayed with gray symbols connected by line segments. The adiabatic frequencies are indicated by line segments only.
\emph{Bottom panel}: Same as above, but  for the best fitting metal-poor model.}
         \label{Fig:echelle}
   \end{figure}

The best fitting model using nonadibatic frequencies, in other words, the model with the lowest $\chi^2\simeq0.79$, matches the observed frequencies with 13 radial and 13 dipole modes. The best fitting model using adiabatic frequencies has $\chi^2\simeq1.0$. The nonadiabatic frequencies provide a better fit than the adiabatic frequencies. The model itself, though, is only slightly different. We list the model fit properties in Tabl.\,\ref{tab:modelinfo}. 
In the top panel of Fig.\,\ref{Fig:echelle} we provide a direct comparison between the observed frequencies and the p-mode frequencies of the best fitting model. The echelle diagram shows that both the adiabatic and nonadibatic model reproduce not only the general pattern of the observed frequencies, but also the detailed structure of the mode sequences. 


\begin{figure}[t]
   \centering
      \includegraphics[width=0.5\textwidth]{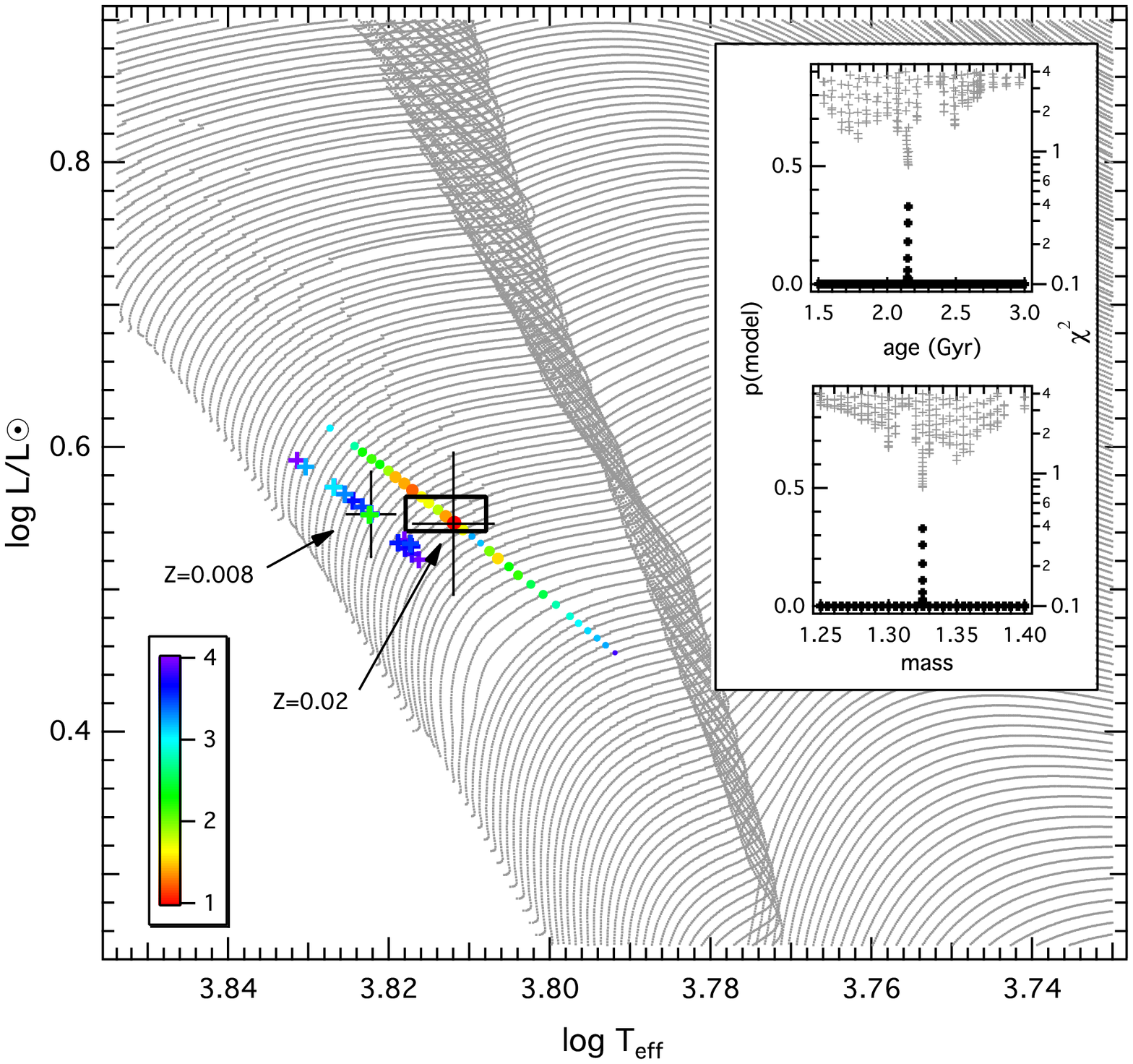}
      \caption{Theoretical H-R diagram showing the uncertainty box of \hd\ and a subset of the solar abundance stellar model grid (light gray dots). The color scale (in the online version only) gives the $\chi^2$ values for the fits of observed and model frequencies, where the scale is limited to values lower than 4.0. Colored circles and crosses correspond to model fits to the solar abundance and metal poor grid, respectively. The grid only is show for the solar abundance models. The large black crosses indicate the positions of the best-fit models. The inserts show the model probability (black filled circles) and the $\chi$ value (gray symbols) distribution for the solar calibrated grid as a function of model age and mass, respectively.}
         \label{Fig:HRD}
   \end{figure}

A theoretical H-R diagram is shown in Fig.\,\ref{Fig:HRD} representing the subset of the stellar model grid used for the pulsation analysis, the error box for the effective temperature and luminosity of \hd , and the best nonadiabatic model fits to the observed frequencies with $\chi^2$ less than 4.0. The adiabatic model fits yield a nearly identical plot. The best fitting nonadiabatic model, located within the error box for \hd , has an effective temperature of T$_\mathrm{eff}$ = 6484\,K, a luminosity of L = 3.52\,L\sun , a radius of R = 1.49\,R\sun, a mass of 1.325\,M\sun , and an age of 2.15\,Gyr (see Tabl.\,\ref{tab:modelinfo}).

Even though our model searches included $l$ = 0, 1, 2, and 3 p modes, our best adiabatic and nonadiabatic model fits do not need to use any $l$ = 2 and 3 modes and, as a consequence, our mode identifications are different from \cite{apo08}. This is not an issue of insufficient frequency resolution. If they could be detected in the data, the $l$ = 2 and 3 mode frequencies form a sequence in the echelle diagram separated from the radial mode sequence by a spacing that is several times greater than the frequency uncertainties. Unlike the $l$ = 0 and 2 modes, the $l$ = 0 and 1 are well separated and nearly independent of each other.

\begin{figure}[t]
   \centering
      \includegraphics[width=0.5\textwidth]{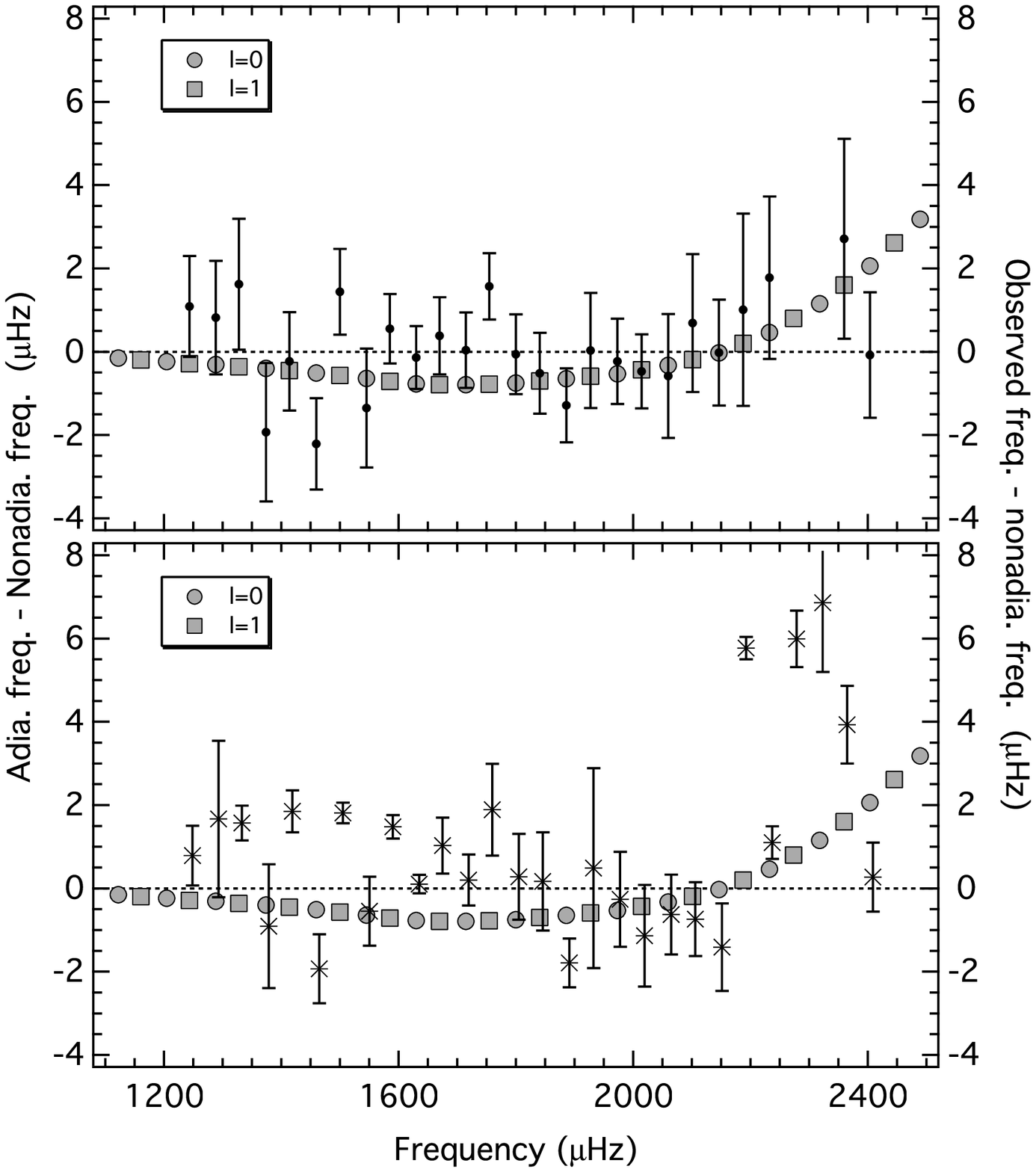}
      \caption{\textit{Top panel:} Difference between the adiabatic and nonadiabatic eigenfrequencies for the $l$ = 0 and 1 modes of the best fitting solar-abundance model and the frequency difference between the observed and nonadiabatic frequencies. Black dots indicate the differences between the observed and the nonadiabatic frequencies with error bars for the observations. \textit{Bottom panel:}  Similar to above except showing the \cite{apo08} frequencies and uncertainties.}
         \label{Fig:adia}
   \end{figure}

If we search our nonadiabatic grid for a best match between the model frequencies given by \cite{apo08} and if we use only those identified by them to come from $l$ = 0 and 1 modes we, find no model with a $\chi^2$ better than 19.6. This model (1.325\,M\sun , 6487\,K) is very close to our best model (see Tabl.\,\ref{tab:modelinfo}) and indicates that -- on average -- the frequency subset derived by \cite{apo08} is similar to our frequency set, which is not really surprising as we are using the same data. The high $\chi^2$ value, however, indicates that their frequency values scatter quite a bit around those predicted by (our) models and the frequency uncertainties are underestimated (by up to a factor of 5) in contrast to our frequencies and uncertainties, derived with a Bayesian approach, and which result in a $\chi^2 = 0.79$ (see Tabl.\,1 and Fig.\,\ref{Fig:adia}). But more importantly, the mode identifications in our analysis are different from that is advocated by \cite{apo08}. If we force our search program to use their frequencies and $l$-value identification, we obtain a $\chi^2 >  45$ and a model temperature difference of more than 1000\,K.

The mass and age of our best model fit depends on the assumed composition of the model grid. We choose primarily for reference purposes a solar composition grid with (Y, Z) = (0.27, 0.02). As discussed in Sect.\,\ref{sec:fun}, our model atmospheres indicate that \hd\ is metal poor compared to the Sun, but with solar C and O abundances! Therefore, a more realistic composition for \hd\ would be closer to (Y, Z) = (0.24, 0.008) assuming normal rates of the Galactic enrichment of helium and metals. Because we do not have a full grid for this specific composition available, we computed a smaller grid of about 10\,000 models centered on the previously best fitting model. 

Repeating the analysis with the lower Z adabatic and nonadiabatic grids, we obtained the echelle diagram for the best fitting model, which is presented in the bottom panel of Fig.\,\ref{Fig:echelle}. Apparently, the low-Z fit is not as good as the solar-Z fit. One must be careful, though, not to conclude that \hd\ has a solar Z, since we do not know the intrinsic model uncertainties exactly. For our Sun, for example, we know that the frequencies of our best solar models differ by up to 0.5\% from the observed frequencies for higher radial orders, i.e., higher frequencies. This discrepancy is attributed to known deficiencies in the modeling of the surface layers. 

With frequency uncertainties in asteroseismic observations approaching $\pm$0.5\mh , it is now essential to consider the effects of the surface layers on the models. For the Sun we know that our inadequate modeling of the surface layers, especially in the region of the superadiabatic layer, results in model frequencies that depart from the observed solar frequencies by more than 10\mh\ at the highest frequencies \citep[see][and references therein]{gue96}.
The model deficiencies have been identified as including uncertain low temperature opacities, inadequate model atmosphere, incomplete inclusion of nonadiabatic effects due to radiation loss and convection, and inadequate mixing length approximation used to model convective energy transport. 

We ourselves are working toward including proper stellar atmosphere models in the model physics and improved structure descriptions in the outer convective envelopes based on 3D numerical simulations of convection \citep{dem99}. But this research is still far from completion. At this time, we can either desensitize our model fitting to the effects of the surface layers or try to estimate their size and note their effect on our model fit conclusions. Because we want to directly address the question of how important and what the influence is of surface effects on our model fits we have chosen the latter.

Regardless, there are several methods in use today to remove the sensitivity of model-fitting to the surface layers that are worth reviewing in our current context. \citet{kje08} use a power-law fit to correct for the effect of the surface layers on the frequencies. The power-law exponent is determined from best model fits to the sun and then the scaling factors are determined from the model fits, themselves, to the observations. The method relies on correctly identifying the observed modes (their $l$ and $n$ values) and assumes that the surface layers scale from the Sun. Although this approach is useful for stars that are known to be like the Sun, we believe it makes too many assumptions about the nature of the surface layers of stars and may cause to misleading results for other stars. 

Roxburgh \citep{rox00,rox01,rox03,rox05} recommends either using ratios of small to large spacings, which are insensitive to surface layers, or computing internal phase shift corrections for the models. The former is easier to implement, but the latter will ultimately yield more information about the location and size of the surface layer effects. The latter method, though, still has to be fully integrated into a $\chi^2$ minimization approach that does not assume $l$ and $n$ identifications. Both approaches do yield superior interior model fits to the observations. They do not teach us what is missing from our models.

To obtain an estimate of the size of the effect of the missing physics in our surface layer modeling, we compared model fits using adiabatic frequencies to fits using nonadiabatic frequencies. The nonadiabatic frequency computation \citep{gue94} accounts for energy gains and losses due to radiation, primarily occurring in the superadiabatic layer of the star. We did not compute the gains and losses caused by convection nor do we include the effects of turbulent pressure on the modes. For the Sun, nonadiabatic radiation effects account for almost half of the discrepancy between observed and modeled frequencies. 

The model-fitting results as represented in the Echelle diagrams of Fig.\,\ref{Fig:echelle} show that the nonadiabatic mode frequencies are close to the adiabatic frequencies, only showing a difference greater than the observational uncertainties above 2200\mh . In terms of $\chi^2$ fits, we do not see a significant difference. The nonadiabatic frequencies are preferred with a slightly lower $\chi^2$ (see Tab.\,\ref{tab:modelinfo}). The best-fit models are also nearly identical. For the current set of observed frequencies below 2200\mh , surface effects, i.e., inadequacies in the modeling of the surface layers, are comparable to the observational uncertainties. Figure\,\ref{Fig:adia} shows this most dramatically. We plot the frequency difference between the adiabatic and nonadiabatic model frequencies versus frequency. We also plot the frequency difference between the observed frequencies and the nonadiabatic frequencies with error bars showing the observed frequency uncertainty. The scatter of the observed minus nonadiabatic frequencies and the error bars are larger than the frequency difference between the adiabatic and nonadiabatic frequencies up to 2200\mh . This suggests that the surface effects are less than the error bars.

The best-fit, nonadiabatic metal-poor model has $\chi^2\simeq2.26$, a mass = 1.205\,M\sun , an age = 2.98\,Gyr, a radius = 1.43\,R\sun , an effective temperature = 6644\,K , and a luminosity = 3.57\,L\sun . The low Z model is located outside the error box of \hd\ in the HRD, which would indicate the need for considerably improved stellar models. The adiabatic model fit is not as good with a $\chi^2 \simeq 4.16$ (see Tabl.\,\ref{tab:modelinfo}).

\subsection{Bayesian mode matching}   \label{ssec:Bayes}

The $\chi^2$-method used so far gives a reasonable result, and one can intuitively ``judge" (also based on the quite different $\chi^2$ values) which of the two fits is better. As noted above, this does not imply that the derived model parameters (mass, age, etc.) are correct since they depend on the assumed model physics. A disadvantage of the $\chi^2$-method is the lack of comparability of different model fits in a statistically satisfying way.
Because we have normalized the $\chi^2$ by assuming the frequencies are independent, which they are not, we can only compare $\chi^2$ fits when the number of frequencies is the same.  In other words, we can identify the model that fits a given set of observations best, but we cannot quantify in terms of a probability how much better that model fit is to another set of frequencies. Similarly, fits from different grids cannot be compared directly, because we have no a priori way of normalizing the $\chi^2$. In the present case, the fit to the solar-calibrated grid appears better than the fit to the metal-poor grid, but we cannot quantify how much better it is. Finally, the mode dependencies on the structure are not completely independent. For example, pairs of modes defining the small spacing have similar eigenfunction shapes in the surface layers, only departing in the deep interior. As a consequence, the $\chi^2$ results normalized by 1/N actually depend slightly on which N modes are matched. In the following we outline our approach to interpret the available information in terms of a probability, based on the Bayesian theorem.

In general, Bayesian techniques have proven to be very successful in comparing observations with models such as \citet{jor05} or \citet{baz08}, which determine some stellar parameters using isochrones and observed values of effective temperature and luminosity. Here, though, we utilize the seismic data to constrain the models.


From the Gaussian distribution of the observational uncertainty it follows that the likelihood that a model frequency $\nu_{m}$ matches an observed frequency $\nu_{o}$ with an uncertainty $\sigma_{o}$ is,
	 \begin{equation}
	 p(\nu_{m}) = \frac{1}{\sigma_{o}\sqrt{2\pi}} \exp \, \biggl ( \frac{-(\nu_{m} - \nu_{o})^2}{2\sigma_{o}^2}\biggr ).
	\label{eq:b1}
	\end{equation}
In fact, we subsequently ignore the normalization factor, $(\sigma_o\sqrt{2\pi})^{-1}$, because it cancels out in the subsequent analysis.
 
As with the $\chi^2$ method, we search our grid of models for model(s) whose frequencies fit the observed frequencies best, but in this case we are looking for the highest probability rather than the lowest $\chi^2$. In the Bayesian approach, though, we can assign an overall probability of the model $M$ matching the observed frequencies with respect to the entire set of models according to Bayes' theorem as
	\begin{equation}
	p(M | D, I) = \frac{p(M | I) \cdot p(D | M, I)}{p(D,I)} ,
	\label{eq:c1}
	\end{equation}
where 
	 \begin{equation}
	p(M | I) = \frac{N_o}{N_{tot}}\mathrm{\,\,\,\,\,\,\,\,} and \\
	p(D | M, I) = \prod_{i=1}^{N_o}  p(\nu_{m,i})
	\label{eq:c2}
	\end{equation}	
are the uniform prior probability for a specific model and the likelihood function, respectively. For the former we use the percentage of the total number of observed frequencies ($N_{tot}$) that are used for the fit ($N_o$). As aresult, the more frequencies are used the higher is the resulting probability.

The likelihood function is the ``and" probability of the individual probabilities of the most credible model frequencies, which were identified in the search for the best model fit. In other words, p(D;M,I) is the probability that $p(\nu_{m,1}) \bigwedge p(\nu_{m,2}) \bigwedge ... \bigwedge p(\nu_{m,N_o})$ fit the corresponding observed frequencies.

The overall model probability is a very sensitive tool because a high probability is only assigned if (almost) all observed frequencies can be matched well. It has to be mentioned that in practice we allow a model frequency to be assigned to more than one observed frequency. This is especially important for example with rotational split modes where several observed modes need to be matched to a single-mode frequency, since the models do not include rotationally split modes. 
The denominator of Eq.\,\ref{eq:c1} is a normalization factor for the specific model probability in the form of
	\begin{equation}
	p(D, I) = \sum_{i=1}^{N_m} p(M_i | I) \cdot p(D|M_i,I).
	\label{eq:c3}
	\end{equation}
Since the uniform priors are the same for all models they cancel in Eq.\,\ref{eq:c1}, which simplifies to
	\begin{equation}
	P = p(M_i | D, I) = \frac{p(D | M_i, I)}{\sum_{j=0}^{N_m} p(D | M_j, I)}.
	\label{eq:c4}
	\end{equation}
The resulting model probability distribution in the model space shows regions of highest probability in contrast to others. This automatically translates into uncertainties in the fundamental parameters of the stellar models by constructing the marginal distribution of the corresponding model parameter. The normalized probability of the most probable model is therefore a measure of how likely this model is with respect to the other models of the specific grid. We stress that it does not tell us how probable the model fit is in an absolute sense. The probability is restricted to the space of the models being considered and their associated physics. 

When comparing solutions of different fits (e.g., for different model grids, or using different sets of observed frequencies) a more robust value for the ``quality of the fit" is the geometric mean of the frequency probabilities (Eq.\,\ref{eq:b1}) of the best fitting model:
	\begin{equation}
	Q = \biggl ( \prod_{i=1}^{N_o}  p(\nu_{m,i})\biggr )^{\frac{1}{N_o}}.
	\label{eq:qual}
	\end{equation}
The best-fit model identified by the Bayesian approach is the same model as that identified by the $\chi^2$ algorithm with the added benefit that the Bayesian formalism provides ranges in which model parameters can be constrained. A full mathematical description and extensive tests of this probabilistic approach will appear separately (Gruberbauer et al., in preperation). As a demonstration we use the Bayesian technique to compare the probabilities of the best fitting models to the solar and low Z (nonadiabatic) grids. The best-fit model to the solar grid has an overall probability of 0.33 with a quality of the fit of $\sim$0.09. 


   \begin{figure}[b]
   \centering
      \includegraphics[width=0.5\textwidth]{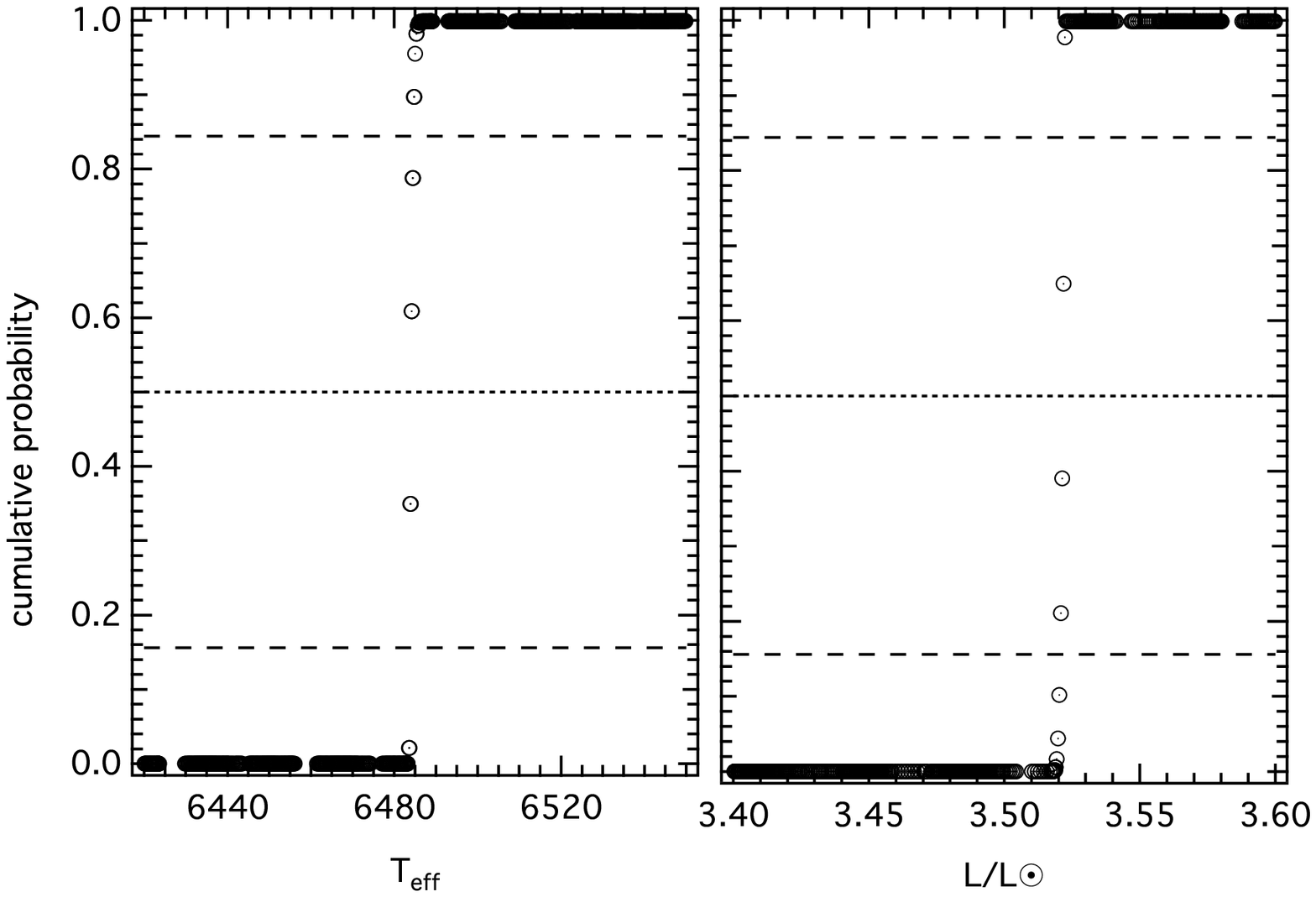}
      \caption{The cumulative probability distribution functions for the effective temperature and luminosity of the solar calibrated grid. The dotted lines correspond to the median and the dashed lines give the $\pm 1 \sigma$ confidence interval.}
         \label{Fig:CD}
   \end{figure}

Using the $\chi^2$-method to find a best fit between model and observed frequencies we found a ridge of models with $\chi^2$ below a given threshold, which are basically located on a contour with a constant large frequency separation where the ridge spans quite a wide range in temperature and luminosity (of course depending on the chosen threshold). The model parameters cannot be constrained in an objective manner.

\begin{table*}[t]
\begin{center}
\caption{Summary of the best fitting solar-abundance and metal-poor adiabatic and nonadiabatic models for the frequencies given by A: \cite{gru08} and B: \cite{apo08}. 
\label{tab:modelinfo}}

\begin{tabular}{lcc|cccccc|cc}
\hline
\hline
\noalign{\smallskip}
Source & (Y, Z)&adia./nonadia.&M & R & Teff & L & log\,g & age & $\chi^2$ &  $Q$\\
            &          &                       &[M\sun ] & [R\sun ]             &[K]     &  [L\sun ]            &[cm$\cdot$s$^{-2}$]&[Gyr]&&\\
\noalign{\smallskip}
\hline
\noalign{\smallskip}
A&(0.27, 0.02) &nonadiabatic& {\bf 1.325} & {\bf 1.49} & {\bf 6484} & {\bf 3.52} & {\bf 4.22} & {\bf 2.15}  & {\bf 0.79} & {\bf 0.090}\\
A &         -        &      adiabatic&        1.325  &       1.49  &        6485  &       3.52  &        4.21  &        2.13   &       1.0  &                0.081 \\
                     
A &(0.24, 0.008)&nonadiabatic& 1.205 & 1.43 & 6644 & 3.57 & 4.22 & 2.98 &  2.26 &  0.043 \\
A &         -          &      adiabatic& 1.205 & 1.43 & 6641 & 3.57 & 4.21 & 2.99 &  4.16 &  0.017 \\
\noalign{\smallskip}
\hline
\noalign{\smallskip}
B&(0.27, 0.02) &nonadiabatic& 1.325 & 1.49 & 6487 & 3.52 & 4.22& 2.14&  19.6 &  1.8$\cdot 10^{-5}$\\
        
 B$^{*}$&      -    &nonadiabatic& 1.600 & 1.53 & 7447 & 6.42 & 4.28 &0.18 &  44.9 & 8.3$\cdot 10^{-11}$\\
 B&(0.24, 0.008)&nonadiabatic& 1.205 & 1.43 & 6645 & 3.57 & 4.22 &2.97 &  27.5 & 3.7$\cdot 10^{-7}$\\
\noalign{\smallskip}
\hline
\noalign{\smallskip}
& &&&1.46$\pm$0.05&6500$\pm$75&3.58$\pm$0.1&4.0$\pm$0.15&&&\\
\noalign{\smallskip}
\hline
\end{tabular}
\end{center}
\end{table*}

As an example, in Fig.\,\ref{Fig:CD} we show the cumulative (i.e., integrated or summed) probability distribution functions of the models as a function of effective temperature and luminosity, respectively. From them, we can derive the most probable model parameters and their uncertainties as, e.g. the median and the $\pm 1 \sigma$ environment. The latter refer to a Gaussian error distribution and can easily be replaced by, e.g. a confidence interval. The cumulative probability distribution functions, however, yields a most probable effective temperature and luminosity of 6484$\pm$1\,K and 3.5205$\pm$0.001\,L\sun , respectively. If the frequency uncertainties were larger, than the high probability would be distributed amongst more models leading to a shallower cumulative probability distribution, and the uncertainties of the model parameters would be larger. The uncertainties are not the uncertainties within which we can determine HD49933's position in the HD-diagram, but they instead correspond to the uncertainties within which we can constrain the model parameters \emph{within a given grid}. The uncertainties appear unrealistically small because we have not included any assumptions about the model uncertainties themselves. 

By applying the Bayesian method to the metal-poor model grid, we find the best model fit (identical to the $\chi^2$ best model) has an overall probability of about 0.39 and a quality of the fit of $\sim$0.043. Interestingly, the overall probability of the best fitting model is higher for the metal-poor grid than for the solar one. One cannot conclude that the metal-poor grid fit is better because the best fitting probability is a measure of the probability of the best-fit model compared to all the other models in its grid and not to the other models' grid. To directly compare the fits of different grids it is necessary to compute the global likelihood, i.e. the sum of the individual model probabilities from each grid. This is done in the following way.

The probability that grid A provides better model fits to a given set of frequencies than grid B (assuming that the models themselves are correct) is given by
	\begin{equation}
	p(A) = \frac{ \frac{1}{N_A} \sum_i p(D|M_{A,i}, I)} { \frac{1}{N_A} \sum_i p(D|M_{A,i}, I)  +   \frac{1}{N_B} \sum_i p(D|M_{B,i}, I)}
	\label{eq:c3}
	\end{equation}
with $N_A$ and $N_B$ the total number of models in grids A and B, respectively. 
Accordingly, we find the probability that the nonadiabatic frequencies of the solar-abundance grid fit the observed frequencies better than the adiabatic frequencies to be about 0.92, which is higher than to the probability of about 0.08 that the adiabatic frequencies provide a better fit than the nonadiabatic frequenices. It becomes even clearer if we compare the different metallicity fits. Whereas the probability that the low Z (nonadiabatic) grid fits the observed frequencies better than the solar Z (nonadiabatic) grid is only about 5.7$\cdot 10^{-9}$, the probability that the solar Z grid fits better than the low Z grid is naturally almost 1.
In other words, of the two grids, the solar Z grid of models is a significantly more likely fit to the observations than the low Z grid of models. The obvious limitation of this conclusion is that it is based entirely on the assumption that the models themselves are correct. This is why a broader range of model parameters and physics needs to be examined before one draw any reasonable conclusions. But our goal here, though, is primarily to demonstrate the potential of the Bayesian approach.

Even though the new approach yields excellent results in the present case, it is very sensitive to the presence of artifacts, i.e., if some of the observed frequencies are not intrinsic to the star, then the Bayesian approach will (correctly) not identify any models as being probable. This is not an issue for the modes we identified for \hd\ in Paper I, but can be a problem for other analyses with lower quality data sets. From Eq.\,\ref{eq:c2} it can be seen that a single observed frequency that does not come close to a model frequency (and consequently has a very low probability of being matched by one of the model frequencies) pushes the overall model probability close to zero. This in turn justifies our very conservative frequency determination described in Paper I. In Gruberbauer et al. (in prep.) we will present a modified version of the algorithm that takes possible artifacts  into account.

\section{Conclusions}		\label{sec:concl}
In Paper I we used a Bayesian approach to determine pulsation frequencies of \hd\ in photometric data obtained by CoRoT. In this paper we have introduced a probability, based on Bayesian methodology, to quantify how closely a model matches the observed pulsation frequency spectrum within uncertainties, compared to other models in a dense and extensive grid of model oscillation spectra. 

We obtained the following results from our analysis.

\begin{enumerate}
\item We first compared the 26 frequencies to model oscillation spectra in our solar composition grid. We identified a best fit with $\chi^2 \simeq$ 0.79 using nonadiabatic frequencies. The luminosity and effective temperature of the best model fit lies within the uncertainty box for \hd\ in the H-R diagram. The mass, age, and radius of the best model fit is 1.325\,M\sun, 2.15\,Gyr, and 1.49\,R\sun .
\item Owing to the uncertain interior metal abundance, we repeated the search with a lower Y and Z grid of models, (Y, Z) = (0.24, 0.008). We identified a best fit with $\chi^2 \simeq$ 2.26. The luminosity and effective temperature of the best-fit model lie outside the uncertainty box in the HR diagram for \hd .
\item We used a Bayesian approach to assign a probability to the model fits and confirmed that the solar Z fit is significantly more probable than the low Z model fit, assuming that the models themselves are valid. 
\item We showed that the mode identification of \citet{apo08} yields significantly higher $\chi^2$ values than our mode identification.
\item We showed that the adiabatic and nonadiatic frequencies yield nearly identical model fits to the observed modes. For the Sun it is known that the nonadiabatic effect accounts for a large portion of the surface layer effect. We therefore conclude that the deficiencies in modeling the stellar surface layers only have weak effects on the present analysis.
\end{enumerate}

A summary of the best model fits to our frequencies (derived in Paper I) and to a subset of frequencies listed by \citet{apo08} is given in Tabl.\,\ref{tab:modelinfo}. In the latter case we use only frequencies which the authors attribute to $l$ = 0 or 1 modes but ignoring their mode identification. Only for case B$^{*}$ do we enforce their mode identification. $\chi^2$ is normalized. Lower $\chi^2$ values indicate a more probable fit. The parameter $Q$ is a probability and refers to the quality of the fit (Eq.\,\ref{eq:qual}). Non-seismically determined fundamental parameters are given in the last row.

The frequency extraction by \citet{apo08} yielded some frequencies that they interpreted as $l$ = 2 modes or rotational split components. Our more conservative analysis does not yield any evidence for frequencies of higher order than $l$ = 1. When we looked for statistically significant modes adjacent to the well-identified modes, we could not find any evidence for additional modes or rotational split components. We suggest that possibly the a priori assumption by \citet{apo08} that the $l$ = 0, 1, and 2 modes have a fixed ratio of heights may have led to spurious $l$ = 2 mode identifications. Likewise, we came to a different mode identification than \cite{apo08}.

Our estimates of the surface abundances of Fe, C, and O reveal that Fe is under abundant compared to the Sun and that C and O have near solar abundances. Owing to this ambiguity we fit the observed frequencies to both a solar composition grid and a low Z grid. The convective envelope mass of the best-fit model in the solar composition grid is $\sim$0.0008\,M\sun . 

We note that the abundances of both helium and heavy elements in HD49933's thin convective envelope will be affected by diffusion processes such as gravitational settling and radiative levitation. Because of computational difficulties we do not follow element diffusion when the envelope thins below 0.1\% of the star's total mass. The mixed metal abundances at the surface make this star an excellent candidate for detailed modeling that includes Y and Z diffusion. A key question we hope to address in a future paper is, whether we can distinguish and isolate the effects of diffusion, metal abundance, helium abundance, and mixing length parameter.

We introduced a Bayesian approach to define a probability that the observed frequencies match a given model spectrum within a grid of model eigenspectra. The model probability allows one to compare distinct model fits with each other, something that the normalized $\chi^2$ values cannot do. 
Although our analysis of \hd\ has yielded results that appear to be more consistent with non-asteroseismic data than the \citet{apo08} results, we do not consider this to be the final word on the subject and fully expect that \hd\ will provide us with new information about Procyon-like stars. We look to future asteroseismic analyses of the CoRoT data to determine the depth of HD\,49933's convective envelope, to determine its helium abundance, to confirm its location in the H-R diagram and phase of evolution, and to test models of Y and Z diffusion. Where seismologists had difficulty trying to interpret the tentatively identified frequencies in Procyon, a star similar to \hd , CoRoT's observations of \hd\ are of such high quality that we will soon begin testing the models themselves.

\begin{acknowledgements}
TK, MG, LF, and WWW are supported by the Austrian Research Promotion Agency (FFG), and the Austrian Science Fund (FWF P17580). TK is also supported by the Canadian Space Agency. DBG acknowledges the support of the Natural Sciences and Engineering Research Council of Canada. \end{acknowledgements}

\bibliographystyle{aa}
\bibliography{11438}

\end{document}